%% file: ms.tex
\documentclass[conference]{IEEEtran}
\IEEEoverridecommandlockouts
\usepackage{cite}
\usepackage{amsmath,amssymb,amsfonts}
\usepackage{algorithmic}
\usepackage{subcaption}
\usepackage{graphicx}
\usepackage{textcomp}
\usepackage{xcolor}
\usepackage{floatrow}
\usepackage{lipsum}
\usepackage[export]{adjustbox}
\usepackage{multirow}
\graphicspath{ {./figs/} }
\def\BibTeX{{\rm B\kern-.05em{\sc i\kern-.025em b}\kern-.08em
    T\kern-.1667em\lower.7ex\hbox{E}\kern-.125emX}}
\begin{document}

\title{DeepAuto: A Hierarchical Deep Learning Framework for Real-Time Prediction in Cellular Networks \\
}



\author{
	\IEEEauthorblockN{
		Abhijeet Bhorkar\IEEEauthorrefmark{1} \thanks{A. Bhorkar and K. Zhang have equal contribution.}} 
	\IEEEauthorblockA{
		\textit{AT\&T Labs} \\
		ab981s@att.com}

	\and

	\IEEEauthorblockN{
		Ke Zhang\IEEEauthorrefmark{1} } 
	\IEEEauthorblockA{
		\textit{AT\&T Labs} \\
		kz3722@att.com}
	\and
	
	\IEEEauthorblockN{Jin Wang} 
	\IEEEauthorblockA{\textit{AT\&T Labs} \\
		jw5458@att.com}
	
}

\maketitle

\begin{abstract}
	
	Accurate real-time forecasting of key performance indicators (KPIs) is an essential requirement for various LTE/5G radio access network (RAN) automation. However, an accurate prediction can be very challenging in large-scale cellular
	environments due to complex spatio-temporal dynamics, network configuration changes and unavailability of real-time network data. 
	
	In this work, we introduce a reusable analytics framework that enables real-time KPI prediction using a hierarchical deep learning architecture. 
	Our 
	approach, namely {\it{DeepAuto}}, stacks multiple long short-term memory (LSTM) networks horizontally to capture instantaneous, periodic and seasonal patterns in KPI time-series. It further merge with feed-forward networks to learn the impact of network configurations and other external factors.

	We validate the approach by predicting two important KPIs, including cell load and radio channel quality, using 
	large-scale real network streaming measurement data from the operator.   
	For cell load prediction, DeepAuto model showed up to 15\% improvement in Root Mean Square Error (RMSE) compared to naive method of using recent measurements for short-term horizon and up to 32\% improvement for longer-term prediction.   
\end{abstract}

\begin{IEEEkeywords}
	Real-time Streaming Measurements, Predictive Analytics, Telecommunication Networks, Hierarchical Deep Learning
\end{IEEEkeywords}

\section{Introduction}

The overall traffic generated by mobile networks continues to accelerate. Telecom operators are expanding capacity by acquiring 
radio spectrum and deploying new base stations; however, at the cost of Capital Expenditure (CAPEX) and operation expenditure (OPEX). Network management automation is a key enabler for dynamic network optimization and  reducing  CAPEX/OPEX costs. 

Accurate prediction of key performance indicators (KPIs) has become increasingly important as it can help telecom operators for better network optimization and network planning. For example, with real-time predictive analytics, network can be intelligently configured  at the right time and at right place. Thus, the radio resources are better utilized with optimizations such as dynamic load balancing/resource allocation, adaptive traffic treatment and adaptive scheduler selection \cite{DBLP:journals/corr/abs-1803-04311}.
Therefore, the use of machine learning, analytics and artificial intelligence
is inevitable for network intelligence. Accordingly, O-RAN alliance is leading the industry to embed intelligence in every layer of RAN architecture for closed loop automation \cite{oran}.

Real-time prediction of traffic/KPIs in wireless networks is however challenging from following perspectives:
\begin{itemize}
	\item Streaming network measurement: a distributed system is needed to collect network measurement data from network elements with strict latency requirement. 
	\item Multi-scale temporal and spatial dependency: recent history captures instant momentum of traffic change; while, periodicity (daily/weekly pattern) and seasonality (monthly/yearly trend) capture global trends. Lack of multi-scale and long-range temporal structure in the model will lead to inaccurate prediction. Furthermore, traffic in different geographical locations could correlate with each other due to user mobility.
	\item Network configuration change: network configurations generally undergo constant changes, which will impact the KPIs and the user's behavior.
	\item External influence: regular traffic patterns can be distorted by external factors such as weather, holiday and local events (e.g., incidents, festival/sport activities, etc.).
\end{itemize}

It is a big challenge to capture all these factors in a single model, largely due to the high-dimensionality of the input/output of the model (a.k.a., the curse of high-dimensionality). In addition to model development, real-time prediction needs significant computation resources due to low latency requirements of prediction.
This requires us to develop an universal model that can be developed and maintained to predict various KPIs for each and every cells in the network. Note that  creating models at the granularity of per cell and per KPI basis is infeasible as it increases latency and computational difficulties for the KPI prediction in production implementation.

In this work, we introduce an efficient and effective solution for real-time traffic prediction, as a major step towards building an 
eco-system that enables proactive network planning and optimization for the next generations of radio access networks. 

Our major contributions are summarized as follows: 
\begin{itemize}
	
	\item We propose a generic hierarchical deep learning framework that predicts various KPIs at the cell level. 
	\item Our model can capture instantaneous, periodic seasonal temporal patterns, spatial patterns, as well as heterogeneous external factors, such as network configurations, day of week, etc.  
	\item We perform extensive experiments, with real-world LTE network streaming measurement data and validate the performance superiority of DeepAuto over traditional supervised learning models, time-series models. 
	\item We propose a real-time prediction framework for RAN optimization. 
\end{itemize}

The remaining of the paper is organized as follows: 
Section \ref{related_work}, discusses related work.
Section \ref{background} provides background and system architecture.
Section \ref{model} proposes our DeepAuto hierarchical deep learning model.
Section \ref{results} provides performance comparison of the proposed approach with the benchmark methods.
Section \ref{conclusion} summarizes our work with our concluding remarks.

\input{relate}

\input{system}

\input{model}
\input{result}


\input{conclusion}

\bibliographystyle{IEEEtran}
\bibliography{IEEEabrv,ms}

\end{document}

%% file: relate.tex
\section{Related Work} \label{related_work}


Time-series models such as Moving Average (MA) or Autoregressive integrated moving average (ARIMA) have been used in \cite{arima_shu}, \cite{arima_zhou} to predict the future traffic load.
MA can predict a single feature based on its own historical data, leaving other impacting factors unconsidered.
Regression models (such as ARIMA, Vector Auto-regression, etc) allow extra features/variables to be included in the model but usually can only handle small input/output dimensions. 
 Random Forest, a supervised machine learning algorithms has been used in \cite{Zhang:2017:TPB:3139958.3140053} to predict traffic load. The tree based ensemble supervised algorithms (such as random forest, gradient boosting machine etc) handle high-dimension input/output but usually ignore the sequential/temporal dependencies among the inputs or are difficult to model due to increased complexity. 
 Traditional methods usually feed all factors indiscriminately, which often leads to a model with huge parameter space and difficult to optimize.

Deep learning has been used in a variety of contexts in mobile networks. 
 \cite{DBLP:journals/corr/abs-1803-04311} provides a survey of all the deep learning applications.
Cell load prediction has been studied in \cite{Jin_infocom, DBLP:journals/corr/abs-1809-00811}. 
\cite{Jin_infocom, DBLP:journals/corr/abs-1809-00811} show that the traffic demand exhibits spatial and temporal
patterns, which help to predict the traffic load.
\cite{Jin_infocom,DBLP:journals/corr/abs-1809-00811}  study the cell load prediction by spatio-temporal analysis on a grid framework using Long short term memory (LSTM). However, this framework is unsuitable when the cell is not placed in regular grid, which usually is the case.  Furthermore, the analysis is not scalable for nationwide coverage due to high training and prediction complexity. 
Even though models using LSTM are good at modeling the sequential dependency, but they fail to capture mixed sequence inputs at various temporal scales.


%% file: system.tex
\section{System Architecture}
\label{background}
\begin{figure}[htbp]
	\centerline{\includegraphics[scale=0.38]{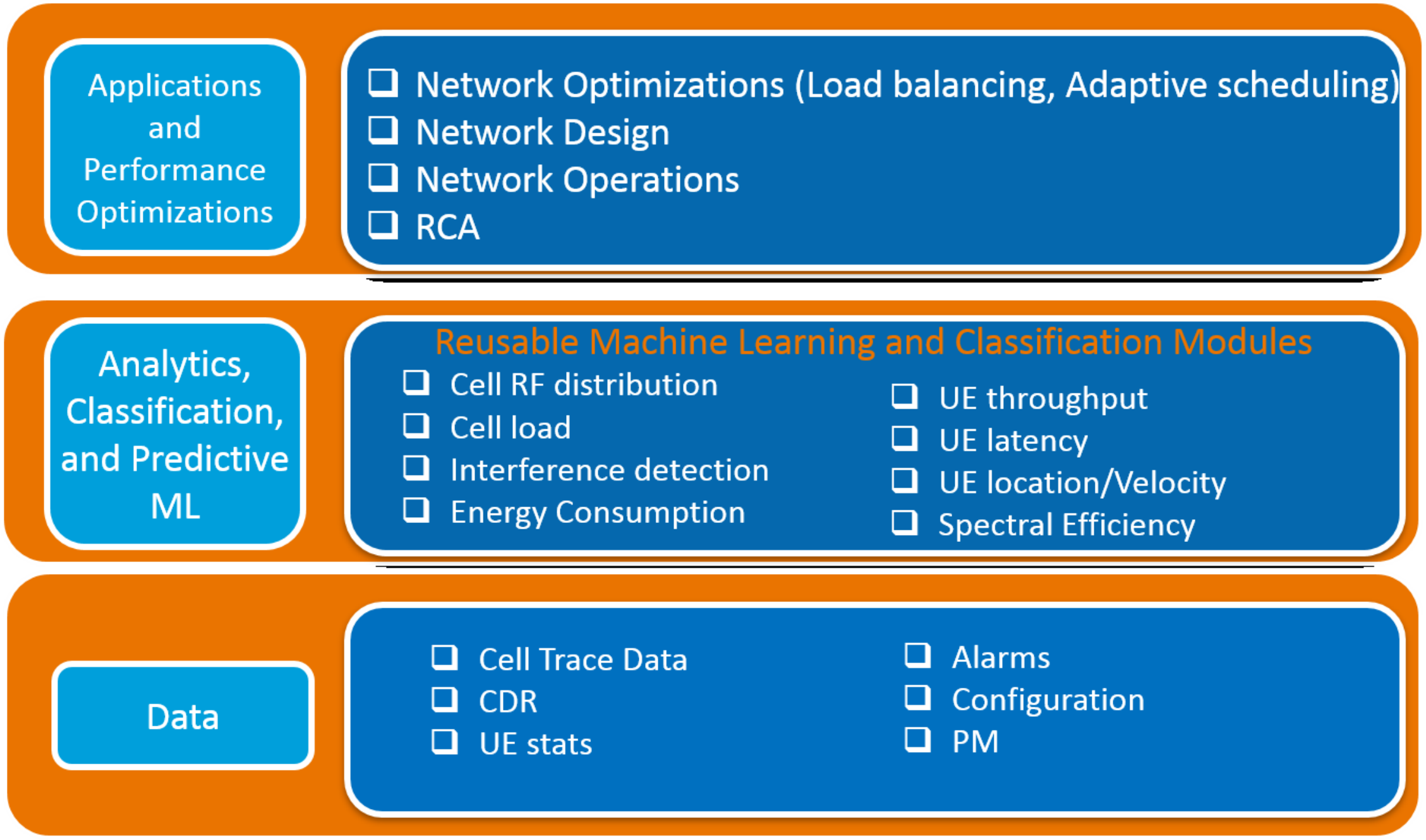}}
	\caption{Various machine learning use cases for network automation.}
	\label{use_cases}
\end{figure}

Fig. \ref{use_cases} provides focus area for LTE/5G RAN applications using artificial intelligence. Network operators over years have developed significant number of data sources from various network elements that may be used to perform classifications and develop predictive algorithms. Potentially these predictions will provide more insights into operations and enable opportunities for performance optimizations. Our aim here is to provide a prediction framework which is extensible across multiple cell level (such as cell load, channel quality) and user (UE) level KPI predictions (such as UE throughput, UE latency, UE BW demand and UE location predictions).

We illustrate our work using most important cell level KPIs including cell load prediction and radio channel quality prediction. Key applications such as adaptive scheduler selection and cell load balancing will be enabled using cell load and channel quality KPI prediction. 

%
\begin{figure}[htbp]
	\centerline{\includegraphics[scale=0.38]{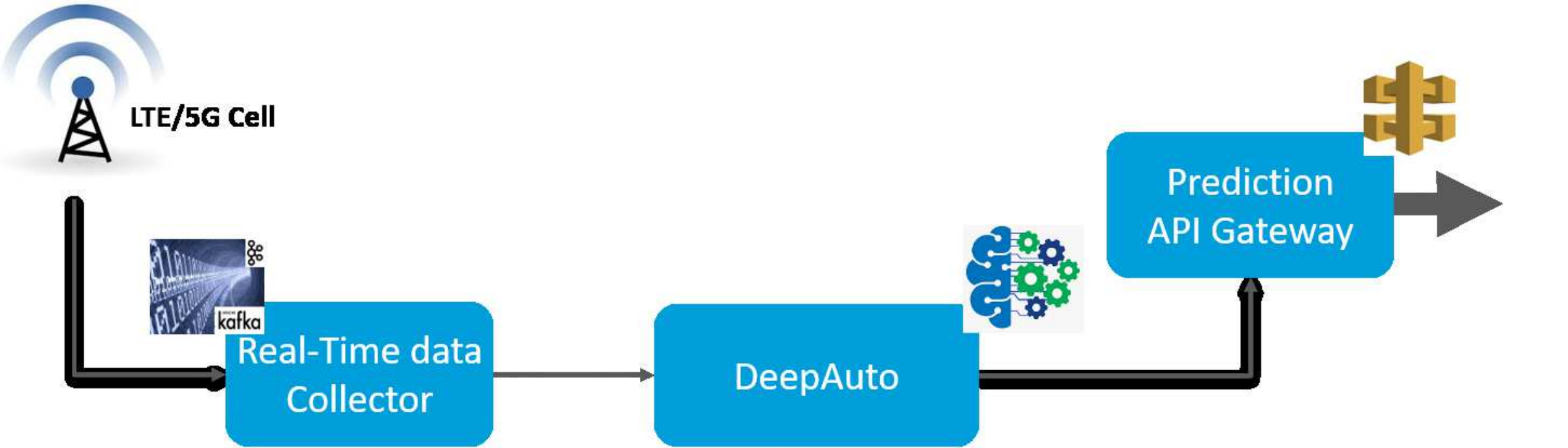}}
	\caption{Real-time prediction system}
	\label{prediction}
\end{figure}

Fig. \ref{prediction}, presents the general framework of our reusable prediction engine. For this prediction, we use real-time data collected from various network elements.  
The real-time collection platform then publishes the data into stream-processing software such as Apache Kafka \cite{Narkhede:2017:KDG:3175825}.
The overall latency of the collection system is order of few seconds. 
Thus, we are able to perform short term prediction with a horizon from a few seconds to a few hours. The data is then consumed by various short term predictors. The predictions are exposed via microservices that provide a unified interface to other downstream applications such as load balancing. 
The real-time data collection architecture is well suited for 5G automation. 

%
%
%

%% file: model.tex
\section{Hierarchical Deep Learning Model}
\label{model}

\begin{figure*}[htbp]
	\centerline{\includegraphics[scale=0.5]{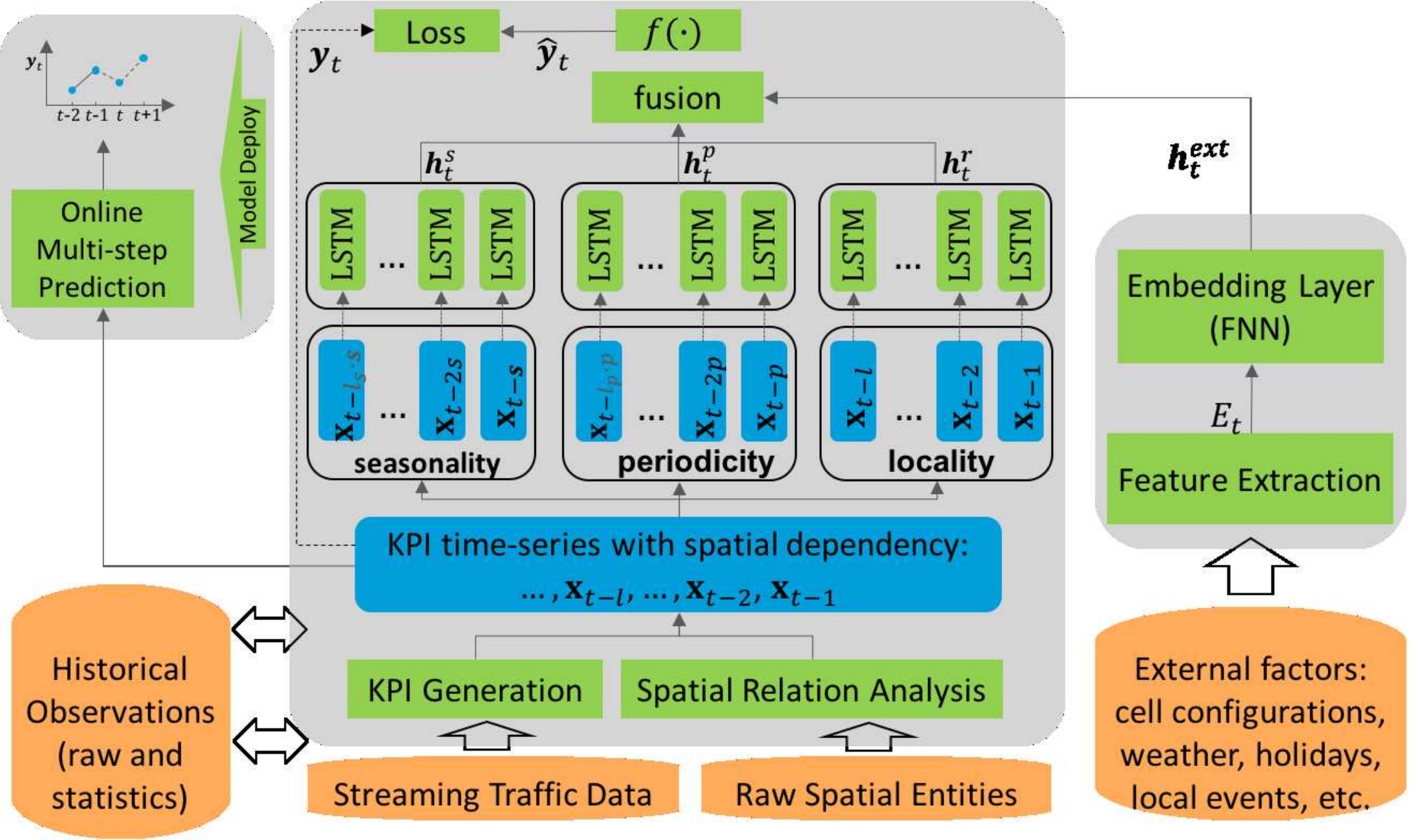}}
	\caption{The structure of the DeepAuto. It has three major components: (i) a hybrid recurrent neural network that is able to capture the patterns of locality, periodicity, and seasonality with different time granularities in time-series data; (ii) an embedding layer using a fully connected layer to learn a representation of external factors; and finally (iii) a fusion layer to merge the effects from different inputs.}
	\label{deepauto}
\end{figure*}

{\bf Problem definition:} the KPI prediction is framed as time-series prediction and in general using Nonlinear Auto-regressive with exogenous inputs (NARX) framework \cite{billings}. 
\begin{equation}
\label{eq:problem}
{\mathbf y}_t = f({\mathbf y}_{t-1}, {\mathbf y}_{t-2}, \cdots,  {\mathbf u}_{t-1}, {\mathbf u}_{t-2}, \cdots) + {\mathbf e}_t,
\end{equation}
where  ${\mathbf y}_t$ represents the vector of variables of interest at time $t$. ${\mathbf u}_t$ is the externally determined variables that have potential impact on the target, and ${\mathbf e}_t$ is the error term. Eqation \ref{eq:problem} can be further written as ${\mathbf y}_t = f({\mathbf x}_{t-1}, {\mathbf x}_{t-2}, \cdots) + {\mathbf e}_t$, where ${\mathbf x}_t$ is the vector concatenation of ${\mathbf y}_t$ and ${\mathbf u}_t$. Given the historical observations $\{\mathbf{x}_0, \mathbf{x}_1, \cdots, \mathbf{x}_{t-1}\}$, the goal is to learn a non-linear function $f(\cdot)$ to predict ${\mathbf y}_t$.  

KPIs in cellular networks can be generated and aggregated with different spatial granularities. 
Depending on the target of interest the spatial granularity can range from a single cell or a base station to a spatial region covering a group of lower-level entities. 
The future dynamics of KPIs often rely heavily on the recent momentum, periodic patterns and seasonal trends. 
A KPI, e.g., cell load, can highly correlate with other KPIs such as number of active user and throughput, as well as KPIs of neighboring cells due to spatial interactions captured by user mobility. 
On the other hand, network configuration updates can have potentially systematic impact on network KPIs, while other external factors, such as weather and local events often lead to abrupt  changes. 
We proposed, {\bf DeepAuto}, a hierarchical deep learning model architecture as shown in Fig. \ref{deepauto}, that can capture heterogeneous temporal, spatial and external factors in a compact and structural way.

The proposed solution ingests streaming network measurement data collected from cellular network. It typically includes periodic samples of cell performance counters and event-driven UE session data.
The model aggregates measurement data for each spatial unit (e.g., cell load for a cell) and calculates a set of KPI as a time-series $\mathbf{x}_t$ with pre-defined time granularity $\Delta t$.
Both raw streaming data and calculated KPI time-series are: a) stored in appropriate database as historical data for model training; and b) fed into deployed model for real-time online prediction.
The model receives input as the recent, near and distant temporal KPIs from given historical observations to model the multi-scale temporal structure of locality, periodicity and seasonality. 
The local input is denoted as $\mathbf{x}^r=\left\{\mathbf{x}_{t-l_r}, \dotsc, \mathbf{x}_{t-2}, \mathbf{x}_{t-1}\right\}$ with $l_r$ timestamps used.
The periodic input is denoted as $\mathbf{x}^p=\left\{\mathbf{x}_{t-k\cdot l_p}, \dotsc, \mathbf{x}_{t-2\cdot l_p}, \mathbf{x}_{t-l_p}\right\}$, where $l_p$ is the period, typically one day.
Likewise, the seasonal part is denoted as $\mathbf{x}^s=\left\{\mathbf{x}_{t-k\cdot l_s}, \dotsc, \mathbf{x}_{t-2\cdot l_s}, \mathbf{x}_{t-l_s}\right\}$, where $l_s$ is a large period capturing the seasonal trend, typically weekly or monthly.

Multiple recurrent neural networks are horizontally stacked to model the multi-scale temporal dependency, that are, $\mathbf{h}_{t}^{r}=f_r(\mathbf{x}^r)$, $\mathbf{h}_{t}^{p}=f_p(\mathbf{x}^p)$ and $\mathbf{h}_{t}^{s}=f_s(\mathbf{x}^s)$. The function $f_*(\cdot)$ represents the recurrent neuron. In particular, we use Long Short-Term Memory (LSTM). 
Unlike classical RNNs, LSTM addresses the problem of long-term dependencies by introducing a purpose-built {\em memory cell} \cite{hochreiter1997long}\cite{rumelhart1986learning} to store information of previous time steps. 
Access to memory cells is guarded by ``input'', ``output'' and ``forget'' gates. Information stored in memory cells is available to the LSTM for a much longer time than in a classical RNN, which allows the model to make more context-aware predictions. 
One typical implementation is via iterating the following composite functions: 

\begin{equation}
\label{eq:memorycell}
  \begin{cases}
    i_t = \sigma(W_{xi}x_t + W_{hi}h_{t-1} + W_{ci}c_{t-1} + b_i) \\
    f_t = \sigma(W_{xf}x_t + W_{hf}h_{t-1} + W_{cf}c_{t-1} + b_f) \\
    z_t = W_{xc}x_t + W_{cf}h_{t-1} + b_c \\
    c_t = f_tc_{t-1} + i_t tanh(z_t)\\ 
    o_t = \sigma(W_{xo}x_t + W_{ho}h_{t-1} + W_{co}c_t + b_o) \\
    h_t = o_t tanh(c_t)
  \end{cases}
\end{equation}
where $\sigma(\cdot)$ is the logistic sigmoid function, $b_*$ are the bias terms and and $i$, $f$, $o$ and $c$ are the {\em input gate}, {\em forget gate}, {\em output gate} and {\em cell} vectors respectively, all of which have the same size as the hidden state vector $h$. 
The weight matrix $W$ indicates the connections between gates, the cell, input and hidden states. 

External features $\mathbf{E}_t$ are extracted from network configurations and streaming external data, such as weather data, weekdays/weekends/holidays. Feed-forward neural networks are applied to learn embeddings of the effect $\mathbf{h}_{t}^{ext}=f_{ext}(\mathbf{E}^t)$.

Finally a fusion layer is designed to aggregate the effects of all factors. Specifically,  $\mathbf{h}_t=[\mathbf{h}_{t}^{r}\Vert\mathbf{h}_{t}^{p}\Vert\mathbf{h}_{t}^{s}\Vert\mathbf{h}_{t}^{ext}]$, where notation $\Vert$ denotes vector concatenation. A final fully-connected layer is applied to predict the target KPI $\hat{\mathbf{y}}_t=f(\mathbf{h}_t)$.

{\bf Spatial Dependency}: 
DeepAuto model allows to include spatial dependency between network entities. It can be determined via traffic interaction and statistical correlation analysis using historical data. A spatial graph is first built where nodes are spatial units and edge weights capture interaction intensity.
For the KPI prediction of single spatial unit (e.g., a cell): top-$k$ neighbors are selected via the ranking of edge weights, e.g., using KPI correlation coefficients. The KPIs of neighbors can be concatenated into vector $\mathbf{x}_t$ as the model input. 

%% file: result.tex
\section{Experiments and Evaluation}
\label{results}
The objective of DeepAuto and the reusable prediction engine is to provide short-term (seconds to few minutes) and mid-term (hours) forecasts of various cell level and UE level KPIs. In this section, we illustrate our model using two important cell level KPI prediction i) cell load prediction ii) channel quality prediction.

\begin{itemize}
\item Cell load prediction: 
 The objective is to predict average cell load a.k.a. Physical Resource Block (PRB) utilization in the next 1 min, 15 min and 1 hour for each cell. PRB utilization for each LTE subframe is the percentage of resource blocks used within each LTE subframe. 
 Average PRB utilization at the cell is computed as the mean of the PRB utilization of each subframe.   
\item Channel quality prediction: 
For channel quality prediction, we use Reference Signal Received Quality distribution (RSRQ), an indicator of interference experienced by the user. 
RSRQ is reported in the radio resource control (RRC) measurement report \cite{3gpp.36.133} with a typical periodicity of 5 seconds. 
Here, our objective is to predict the aggregate RSRQ distribution over next 5 minutes. 
\end{itemize}
 
In the experiments that follows we have provided detailed evaluation of our framework for cell load prediction objective. Finally, we briefly provide results of the channel quality prediction. 

\subsection{Performance Results for cell load prediction}
 
We perform our experiments using different datasets in two phases. 
In the first phase we characterize and show superior performance of DeepAuto against various baseline algorithms. In the second phase, we build a real-time production grade prediction model using a large scale dataset and evaluate performance of future predictions against various metrics. 

i) {\it {Batched PM counters data:}} In the first phase, we collected Performance Measurements (PMs) counters from eNB across the nation aggregated at 15 minutes intervals \cite{3gpp.32.425}. We collected 3 months of data from April 2018-June 2018 for nearly 1.5k cells within the same geographical area (corresponding to one spatial cluster) and about 1M records collected at interval of 15 min.
The amount of raw data collected is around 441 MB (compressed).   

ii) {\it {real-time streaming data:}} We collected real-time Cell Traffic Recordings (CTR) \cite{3gpp.32.423} from various network management system (NMS) across the nation from nearly 500k cells aggregated at an internal of 1 minutes. The data volume collected over 14 days during July 2018 is over 400 GB (compressed) in size. Compared to PM counters, real-time streaming data contains significant number of missing values. Thus compared to CTR data, PM data is more reliable for prediction; however, it incurs additional collection latency. For PM and real-time streaming data, missing values are filled by using linear interpolation. 



\subsubsection{Evaluation (Phase 1)}

\begin{figure}[htbp]
	{\includegraphics[scale=0.5]{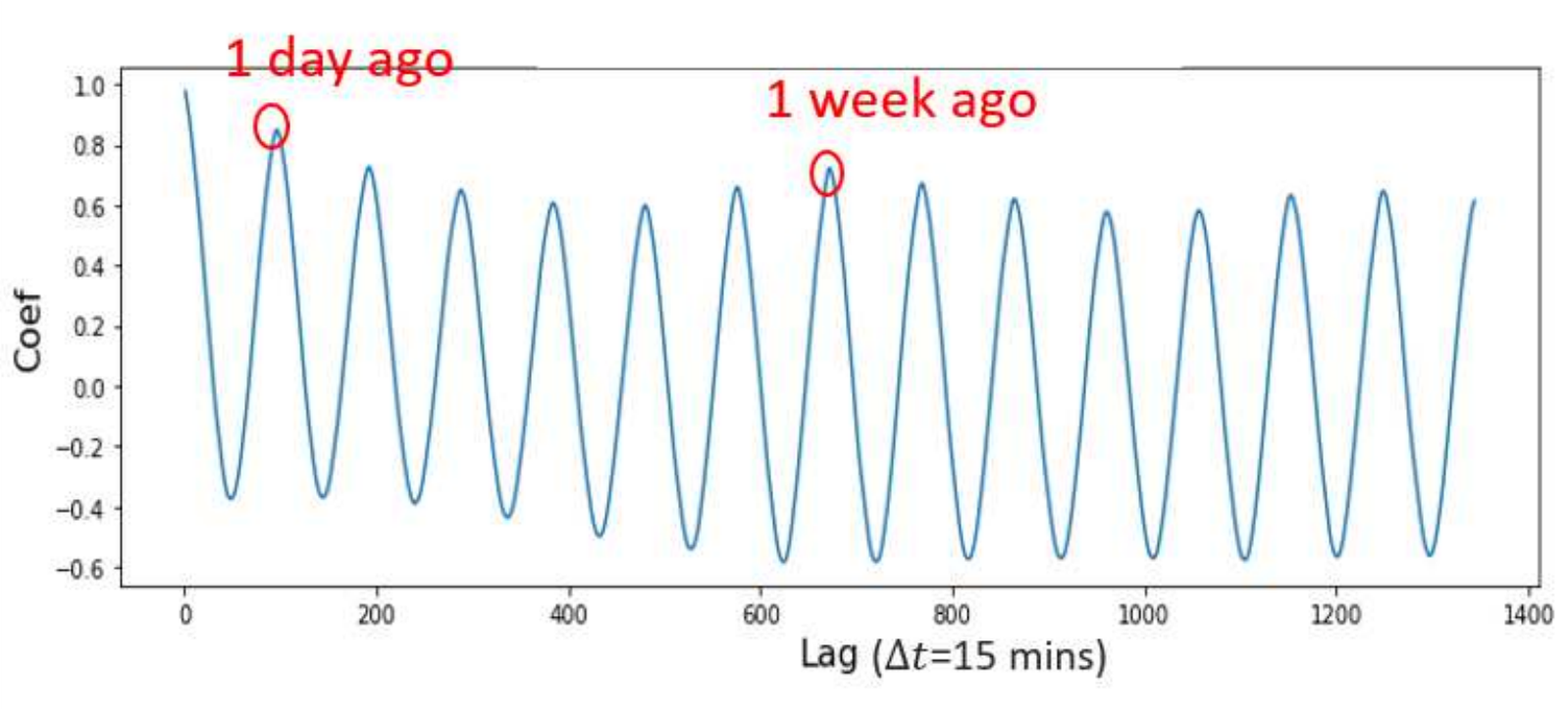}}
	\caption{Auto-correlation of cell load.}
	\label{autocorellationcellload}
\end{figure}

\begin{table}[!htb]
	\begin{tabular}{|c|c|}
		\hline 
		\bf Feature & \bf RMSE  \\
		\hline 
		$l_r$=5, $l_p$=0, $l_s$=0 &  0.0675  \\ 
		\hline 
		$l_r$=10, $l_p$=0, $l_s$=0 &  0.067  \\ 
		\hline 
		$l_r$=20, $l_p$=0, $l_s$=0 &  0.0668  \\ 
		\hline 
		$l_r$=20, $l_p$=1, $l_s$=0 &  0.064  \\ 
		\hline 
		$l_r$=20, $l_p$=2, $l_s$=0, external features &  0.0628  \\ 
		\hline 
	\end{tabular}
	\caption{Effect on locality, periodicity and seasonality on cell load prediction.} 
	\label{tab:seasonalitycellload} 
\end{table}

Next, we present the exploratory experiments conducted using PM counters dataset. 
In order to examine the existence of long-term and/or short-term repetitive patterns in cell load, we plot auto-correlation of cell load in Fig. \ref{autocorellationcellload} for a randomly selected cell. It depicts that correlational is high for 1 day and 1 week. This implies importance of inclusion of periodic and seasonal patterns.  

{\bf{Training DeepAuto model}}:
We use modified mean square error (MMSE) as a loss function during training phase.
\begin{equation}
MMSE = \frac{1}{Kn}\sum_{j=0}^{K-1}\sum_{i=0}^{n-1}\exp^{-\alpha(1-y_{ij})}(y_{ij}-\hat{y}_{ij})^2, 
\end{equation}
where $y_{ij}$ is the true value, $\hat{y}_{ij}$ is predicted value for $i^{th}$ example and $j^{th}$ dimension. Bias $\alpha$ is added to give more importance to those critical examples, e.g., overloaded cell utilization, to achieve a better prediction for critical load. 


The temporal feature set for a cell at time $t$ includes cell load and  number of UEs at cell at time $t$. 
Additional external features $\mathbf{E}_t$ at time $t$ include day of the week, hour of the day, cell configuration details such as band, power and bandwidth. 
During training, data is split in train, validation and test set while maintaining the temporal order of observations in the ratio of 4:1:1 respectively.
As the machine learning models are sensitive to the scale of the inputs, the data are normalized into the range [0, 1] by using feature scaling. 
DeepAuto accuracy is improved by hyper-parameter search. First, we perform a search over parameters $l_r$, $l_p$ and $l_s$ for local, periodic and seasonal trends. 
Table. \ref{tab:seasonalitycellload} shows the performance of DeepAuto for 1 step prediction (15 min horizon) as we optimize temporal features selection by varying the temporal parameters. As expected, including periodic and seasonal pattern improved the accuracy of the results. 
The performance is improved by optimizing learning rate (${lr}$) and batch size. Finally $\alpha$ parameter for the loss function is optimized. 
After hyper-parameter search we use batch size of 1024, $\alpha=4$ and $lr=0.005$.

%


\begin{table}[!htb]
	\begin{tabular}{|c|c|c|c|c|}
		\hline 
		\bf Algorithm & \bf Horizon & \bf RMSE & \bf MAE & \bf MAPE \\
		\hline 
		DeepAuto & 15 min & 0.0628 & 0.0425 & 12.5 \\ 
		\hline 
		Naive  & 15 min & 0.074 & 0.053 & 18.2 \\ 
		\hline
		Random Forest & 15 min & 0.0642 & 0.0432 & 13.1 \\ 
		\hline 
		XGBoost  & 15 min & 0.0638  & 0.0431 & 12.9 \\ 
		\hline
		\hline
		DeepAuto & 120 min & 0.094 & 0.065 & 19.9 \\ 
		\hline 
		Naive  & 120 min & 0.140 & 0.09 & 27.0\\ 
		\hline
		Random Forest & 120 min & 0.098  & 0.067  & 21.1 \\ 
		\hline 
		XGBoost  & 120 min &  0.098  & 0.067  & 20.9 \\ 
		\hline 
	\end{tabular}
	\caption{DeepAuto performance comparison using PM counters.} \label{tab:deepautoresultspm} 
\end{table}

Next, we provide comparison of DeepAuto model against various baseline algorithms. 
Details of the baseline algorithms used in comparison are presented below: 
\begin{itemize}
	\item Naive: In this method, prediction at time $\hat{y}_{t} = y_{t-1}$.
	\item Random Forest: We used random forest model implemented via H2O.ai \cite{h2o_Java_software}. The prediction result is optimized by varying the number of trees in \{50, 100, 200\}, splits rate at each node in \{0,8, 1.0\} and depth of the tree in  \{6, 10, 15\}. 
	\item XGBoost: In this method, we used XGBoost from H2O.ai \cite{h2o_Java_software}. The prediction result is optimized using the same parameter set as mentioned in Random Forest.  
\end{itemize}
For comparison between various baseline algorithms, we use the PM counter batch data source as we were unable to train baselines models with CTR data due to sheer amount of volume. 
For fair comparison, we use the same feature set including temporal and external features for Random Forest and XGBoost as that of DeepAuto.\footnote{Random-forest and XGBoost algorithms in general require the complete training data-set to be loaded into memory for fair comparison. It is usually not feasible to make use them for large scale training.}

Table~\ref{tab:deepautoresultspm} compares the performance of DeepAuto under metrics including Mean Absolute Error (MAE),  Mean Absolute Percentage Error (MAPE) and the Root Mean Square Error (RMSE) with the horizon of 15 min and 120 minutes. 
For MAPE we only consider cells with cell load greater than a threshold of 70\% to reduce the bias for low load cells, where 70\% is the sum of mean and standard deviation of the cell load in the training dataset. DeepAuto performs superior compared to other baseline methods in all metrics considered. 
For 15 min horizon,  
DeepAuto showed upto 15\% improvement in RMSE
compared to naive method, 2.5 \% improvement compared to random forest and  
1.5\% improvement compared to XGBoost.  
For longer horizon of 120 mins, DeepAuto showed upto 32\% reduction in RMSE compared to naive method, 4\% improvement over XGBoost. We observe that the performance improvement of DeepAuto over other method improves with longer horizon.

\subsubsection{Evaluation (Phase 2)}
After initial investigation and validating superiority of DeepAuto, we developed a production grade model for a large scale network deployment. 
During phase 2, we utilize real-time streaming data source with a latency of 1 min.   
Similar to phase 1, we optimize DeepAuto accuracy by optimizing hyper-parameters. We build a model for each of the network management system (NMS) where the real-time data is first received. 
Prediction phase uses real-time Apache Kafka \cite{Narkhede:2017:KDG:3175825} feed from nationwide eNBs.
The prediction engine runs at every regional center close to each of the network management system (NMS) to reduce prediction latency. 
The model generated from the training phase is used to predict the future cells loads for next 1 min, 15 min and 1 hour at a granularity of 1 min. The results are then fetched by various micro-services as needed to cater to various applications. The trained model is regularly updated to capture any tending traffic changes not captured by the model. 

\begin{table}[!htb]
	\begin{tabular}{|c|c|c|c|}
		\hline
		Metric & \multicolumn{3}{c|}{Horizon} \\
		\hline 
		\bf  & \bf 1 min & \bf 15 min & \bf 1 hour\\
		\hline 
		RMSE &  0.083 & 0.066 & 0.067 \\ 
		\hline 
		MAE  &  0.053 & 0.043 &0.044 \\ 
		\hline
		MAPE  &  14.1 & 12.0  & 13.04 \\ 
		\hline 
	\end{tabular}
	\caption{DeepAuto performance results at one of the NMS.} \label{tab:deepautoresults} 
\end{table}
Table \ref{tab:deepautoresults} describes the performance of DeepAuto under various metrics MAE, MAPE and RMSE while predicting cell load for future horizons including next 1 min, next 15 min average cell load and next 1 hour average cell load. Here, we have used average cell load for prediction instead of instantaneous cell load due to noisy nature of CTR data and possibly unreliable predictions. For MAPE we only consider high load cells where cell load is greater than 60\% to reduce the bias of low load cells.
Note that 1 min data is noisy due to real-time nature of data and presence of various missing data points compared to the batched data source. 
Even though DeepAuto allows to exploit spatial dependency, maintaining model for each cluster is restrictive in production. Furthermore, our analysis observed that the use of spatial relationship did not always help to improve the performance. Hence, we decided to deploy a single global model for each of the NMS and maintain the latency requirement with minimal loss of accuracy. 

\subsection{Performance Results for channel quality prediction}

We utilize DeepAuto framework for predicting the RSRQ distribution. The  RSRQ values from the UE are reported at every 5 seconds interval within a range from 0 to 34. To make our analysis more tractable, we group the RSRQ values for each cell by timestamp at 5 min interval. 
The objective is to predict RSRQ probability distribution function (PDF) in the next 5 mins for each cell. We use real-time streaming data source with a volume of about 386 MB (compressed) from 1.5k cells.
We use Kullback-Leibler (KL) divergence as a loss metric for comparing true and predicted distribution. The loss function used during training and testing is given by:
\begin{equation}
L = \frac{1}{n} \sum_i D_i(P||Q),
\end{equation}
where $L$ is the KL divergence, $n$ is the number of LTE cells in the  dataset and $ D_i(P||Q)$  is the KL divergence calculated at cell $i$ as:
\begin{equation}
D_i(P||Q) = - \sum_x p(x) \log q(x)  + \sum_x p(x) \log p(x),
\end{equation}
where $p$ is the actual PDF and $q$ is the predicted PDF. 
Similar to the cell prediction, we use external features such as cell configuration,  day of week, hour of day and minute of day. 


\begin{table}[!htb]
	\begin{tabular}{|c|c|}
		\hline 
		\bf Feature & \bf KL divergence  \\
		\hline 
		$l_r$=5, $l_p$=0, $l_s$=0 &  0.038  \\ 
		\hline 
		$l_r$=10, $l_p$=0, $l_s$=0 &  0.0365  \\ 
		\hline 
		$l_r$=20, $l_p$=0, $l_s$=0 &  0.036  \\ 
		\hline 
		$l_r$=25, $l_p$=1, $l_s$=1 &  0.036  \\ 
		\hline 
		$l_r$=25, $l_p$=1, external features &  0.0353  \\ 
		\hline 
	\end{tabular}
	\caption{Effect on locality, periodicity and seasonality for channel quality prediction.} 
	\label{tab:seasonalityrfprediction} 
\end{table}

Table. \ref{tab:seasonalityrfprediction} compares different combination of features, locality, periodicity, seasonality, and externality. As expected, including additional temporal improves prediction accuracy.  
Note that traditional statistical/machine learning methods seem unsuitable for this problem. Thus, baseline performance is not provided for other methods such as ARIMA, random-forest. Naive method of using $y_{t-1}$ as prediction resulted in KL divergence of 0.14 while DeepAuto  achieved a KL divergence value of 0.0353 (over 75\% improvement compared to naive method). 



%% file: conclusion.tex
\section{Conclusion}
\label{conclusion}
Accurate forecasting of RAN KPIs represents an essential part LTE/5G RAN automation. We provided an unified, efficient and effective traffic prediction architecture that predicts various RAN KPIs in real time. 
We presented the prediction model {\bf{DeepAuto}}, hierarchical deep learning framework, that constructively captures spatial, temporal and external factors, as well as network configuration changes in a scalable manner. We validated our framework using two KPI prediction: cell load prediction and channel quality prediction. We showed that DeepAuto is able to forecast accurately over both short term to medium term time horizon.
Specifically, DeepAuto reduced the prediction error by upto 15\% in RMSE for short term cell load prediction, 32\% gain in long term cell load prediction and 75\% improvement in KL divergence for channel prediction compared to the naive method of using recent measurements.